# *Dependence of Zinc Oxide Thin Films Properties on Filtered Vacuum Arc Deposition Parameters*

*T. David*[*1], *S. Goldsmith*[1], *and R.L. Boxman*[2]

[1] Tel Aviv University, Raymond and Beverly Sackler Faculty of Exact Sciences, School of Physics and Astronomy

[2] Tel Aviv University, Iby and Aladar Fleischman Faculty of Engineering, Department of Interdisciplinary Studies.

Electrical Discharge and Plasma Laboratory,
Tel Aviv University, Tel Aviv 69978, ISRAEL

## *Abstract*

The micro-properties (structure and composition), and macro-properties (electrical and optical properties) of zinc oxide (ZnO) thin films deposited on glass substrates using a filtered vacuum arc deposition (FVAD) system were investigated as a function of oxygen pressure (0.37 – 0.5 Pa) and arc current (100 – 300 A). The films were polycrystalline, and the crystal plane orientation varied with the oxygen pressure and arc current, tending to be aligned parallel to the c-axis. The sizes of the crystallite grains were 10 - 35 nm. The films were found to be compressively stressed, with stress in the range of -2.5 – 0 GPa. The stress in any sample decreased as function of arc current, however, its dependence on the pressure itself also depended on the applied arc current. The compressive stress in samples deposited with arc current in the range 100 ⁻ 150 A, decreased with the pressure from -2.5 to -1.5 GPa (0.37 – 0.5 Pa), whereas it increased with the oxygen pressure in samples deposited with arc current 200 to 300 A. The compressive stress in all samples deposited with the highest oxygen pressure (0.51 Pa) was in a relatively narrow range -2.1 to -1.7 GPa. Film composition, determined by X-ray photoelectron spectroscopy (XPS), depended weakly on the deposition parameters. All samples had zinc excess, with typical oxygen to zinc atomic concentration ratio 0.7 – 0.8. Film thickness, in the range of 80 – 780 nm, depended linearly on both deposition parameters.

---

[*] Author to whom correspondence should be addressed. Current email: taldavid@bgu.ac.il

The electrical resistivity ($\rho$) of the films was in the range of $(1-5) \cdot 10^{-4}$ $\Omega$.m, depending weakly on the deposition parameters. The electrical resistivity of the films with larger grain size was higher than that of films with smaller grains, whereas it increased with film stress. The optical transmission of the films, expressed by the extinction coefficient, depended strongly on both deposition parameters (arc current and oxygen pressure). The lowest extinction was obtained with films deposited with higher-pressure (P $\geq$ 0.5 Pa) and lower arc-current (I $\leq$ 200 A). The lowest extinction coefficient was ~$4 \cdot 10^{-4}$ nm$^{-1}$ in the visible and the near-IR range of the spectrum. Films with larger grain size and lower stress had relatively larger extinction coefficient (~$8 \cdot 10^{-3}$ nm$^{-1}$).

Keywords: Filtered Vacuum Arc Deposition (FVAD), Zinc Oxide (ZnO), Elastic Stress, oxygen-to-zinc ratio.

*1. Introduction*

Zinc oxide (ZnO) is a transparent conducting oxide (TCO), which has recently been studied extensively. It is a II-VI semiconductor, mostly n-type, with a wide band gap of ~3.3 eV, that could be obtained with resistivities as low as $10^{-6}$ Ω·m. It is a candidate material for use as a gas sensor, in electronic displays, in the fabrication of blue light emitting diodes (LEDs), in surface acoustic wave (SAW) devices, and more.

ZnO films have been deposited using many methods of deposition techniques. These include various sputtering techniques [1-5], chemical vapor deposition (CVD) [6-8], molecular beam epitaxy (MBE) [9,10], pulsed laser deposition (PLD) [11], sol-gel [12], filtered vacuum arc (FVA) [13-21], and more. Relatively only few reports on the properties of ZnO films deposited with FVA are found in the literature, including a recent publication on this subject by the present authors [22], in comparison to the number of reports where other deposition methods were used, although the FVA deposition method is characterized by a larger deposition rate.

The effects of substrate temperature, substrate bias during deposition, post-deposition annealing in various atmospheres, or doping with various elements, e.g., N, Al and Sb, on ZnO film resistivity, structure, and optical transmission were studied extensively. Most of the reports are elective, in the sense that they focus on a relatively narrow range of deposition parameters or film properties. The number of comprehensive studies of as-deposited undoped ZnO films, deposited by use of vacuum arc deposition systems at room temperature, is even smaller [19,22].

In this paper we report on such investigation, in which film properties are determined as a function of the two basic vacuum arc deposition system parameters: the background oxygen pressure, and the arc current, and the interrelation between the micro and macro properties of the film is assessed. Arc current and oxygen pressure are two "knobs" that can be systematically controlled to determine the characteristics of the deposited ZnO films. The upper and lower limits of the pressure were determined by the requirement that the film would be both conducting and transparent; the upper and lower limits of the current were determined by the current source and by arc stability, respectively.

## 2. Experimental Apparatus and Procedure

The deposition system was previously described in detail [22-24]. It is comprised of a plasma gun with a Zn cathode, and a quarter-torus magnetic macroparticle filter attached to the deposition chamber. Oxygen was injected in the vicinity of the substrate by a controlled valve, and the pressure is kept constant during the deposition by a computerized controller. The substrates were made of 25x75 mm microscope glass slides, which were coated by a non-uniform ZnO film. The analysis was performed on the central uniform section of the sample, 20x20 mm, which was measured to have uniform film thickness to within 10%. [22] The substrates were not heated or biased during deposition, and were not annealed or otherwise treated after deposition.

Based on past work, an assembly of film samples was prepared by running the arcs at oxygen background pressure in the range 0.373 – 0.506 Pa (varied in steps of 0.026 Pa) and the arc current varied in the range 100 – 300 A (mostly in steps of 50 A).

Film composition, on the surface and in the bulk, was determined by X-ray photoelectron spectroscopy (XPS) (studying the oxygen O(1s) peak at 530.5 eV, and the zinc Zn(2p3) peak at 1021.4 eV), using a PHI scanning 5600 AES/XPS multi-technique system. Bulk composition was obtained by sputtering with $Ar^+$ ions a hole through the film, combined with the AES/XPS analysis. Film structure was analyzed by X-ray diffraction (XRD) using a Scintag X-ray diffractometer equipped with a Cu anode (Cu K$\alpha$, $\lambda$=0.1541 nm).

Film thickness was determined by counting the interference fringes formed on the sample. This was occasionally verified by a profilometer, and with mass gains data. The deposition time was kept constant at 60 s, and the thickness depended on the deposition parameters, as will be discussed below.

The electrical sheet resistance was measured using a two-point measurement method by contacting two Cu adhesive tapes attached at the ends of the central uniform section. These measurements were occasionally verified with a 4-point probe.

The film optical transmission was measured with a spectrophotometry, using a Minuteman MV-305 monochromator equipped with a calibrated photomultiplier. This device had limitations that did not enable measurements in the UV; hence no direct data on the band-gap is presented.

## 3. Results and Discussion

### 3.1 Micro-Properties analyses

*3.1.1 General observations from Structural analysis*

Typical XRD spectra of samples deposited with oxygen pressure of 0.426 Pa and arc current 100 – 300 A are presented in Figure 1. These films, like all other films, had polycrystalline hexagonal wurtzite structure, with dominant c-axis orientation, as indicated by the dominant (002) reflection intensity. However, as shown in Fig. 1, in a single case, the (110) reflection was the strongest on the film deposited with 250 A. The intensity of most observed X-ray reflections increased first with arc current, except for the (100) reflection, whose intensity decreased on the films deposited with arc current greater than 200 A or 250 A, depending on the oxygen pressure. The decrease in the intensity of the (100) reflection was observed in films deposited at 200 A and at lower pressure, in the range (0.38 – 0.43 Pa), while in case of films deposited at 0.51 Pa it was observed on films deposited with 250 A. The width and position of the dominant (002) reflection were further analyzed to derive average grain size and internal stress. It was also noted that in most cases the (100) reflection peak intensity was much lower than that of the (002) reflection.

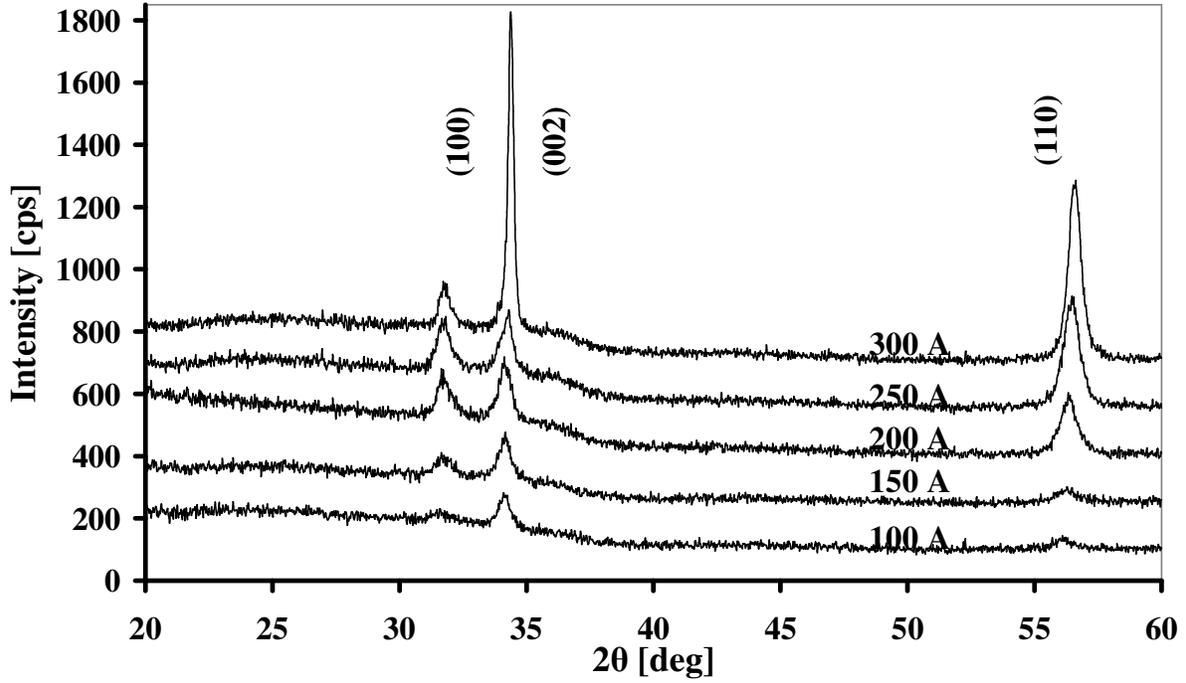

**Figure 1:** X-ray diffractograms of ZnO films deposited with 0.426 Pa oxygen pressure (arc currents = 100 A, 150 A, 200 A, 250 A, and 300 A). The diffractograms were shifted vertically to facilitate the viewing. The base line of each diffractogram is at zero intensity.

*3.1.2 The (002) peak position and elastic stress.*

The elastic stress ($\sigma$) of the films can be determined from Hoffman's relation [25]:

$$\sigma = \frac{2C_{13}^2 - C_{33}(C_{11}+C_{12})}{2C_{13}} \frac{(c-c_0)}{c_0}$$

In this expression the coefficients $C_{ij}$ are the elastic stiffness constants (values taken here are of single crystal ZnO [26, 27], and $c$, $c_0$ are the measured and stress-free c-axis lattice constants, respectively. The lattice constant was deduced from the (002) reflection peak by computing the lattice spacing using Bragg's law [28]. Using the diffractometer wavelength Cu $K_\alpha$, $c_o$=0.5206 nm [29].

A plot of the film elastic stress against arc current, with oxygen pressure as a parameter, is shown in Figure 2. Except for one case, the stress was negative (compressive), depending on the oxygen pressure, and independent of the current below a certain transition current, 200 –250 A. In most cases, film stress in samples deposited with current greater than 200 A decreased significantly as function of the current. In the case of a film deposited with a current of 250 A, and a pressure of 0.37 Pa, it even changed from negative stress to weak positive tensile stress. The variation of the stress as function of the current was moderate in films deposited at lower pressure.

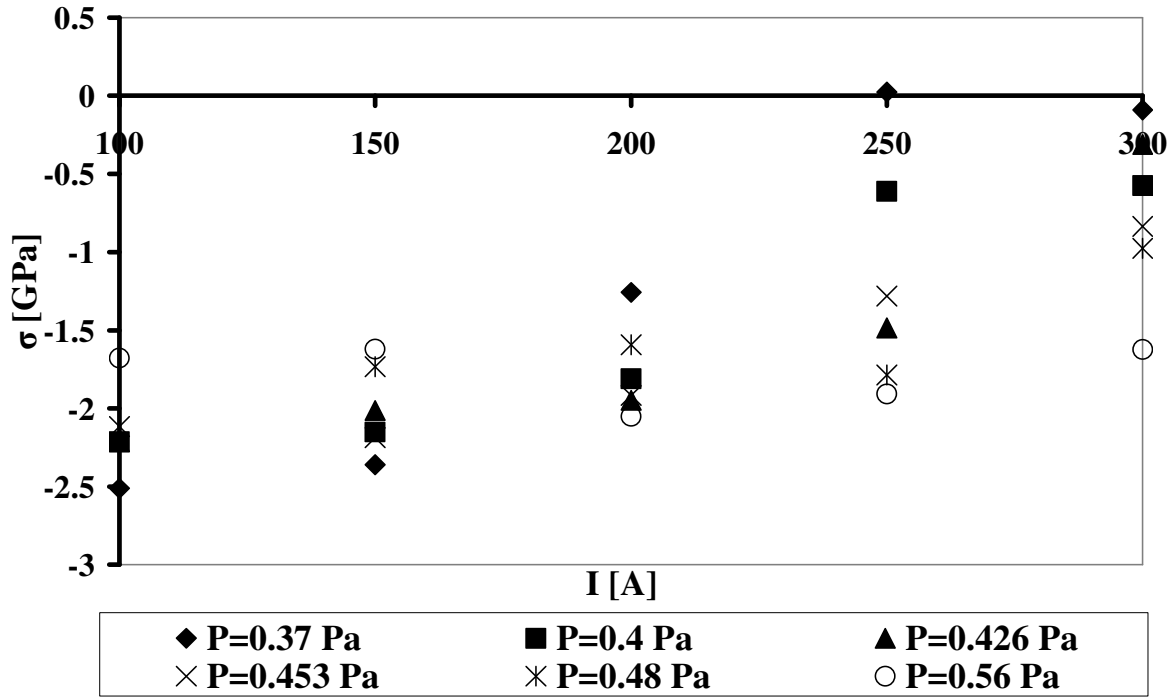

**Figure 2:** Plot of ZnO film elastic stress σ (GPa) vs. the arc current I (A), where the oxygen pressure range was 0.37 – 0.51 Pa.

*3.1.3 The (002) peak intensity*

Peak intensity of the X-ray diffraction reflections is determined by the crystalline grain size and structure, axis orientation, and could also be affected by film thickness. No correlation was established in this study between diffractions reflection intensity and film thickness. As the (002) reflection is the dominant reflection feature, the dominant axis is the wurtzite c-axis. The plots of the (002) reflection peak intensity, as function of the current, where the pressure is a parameter, are presented in Fig. 3. The data indicates weak dependence of the X-ray reflection intensity on arc currents below 250 A. It should be noted that the (002) reflection intensity increased markedly when arc current is greater than 250 A for P=0.37, 0.46 Pa.

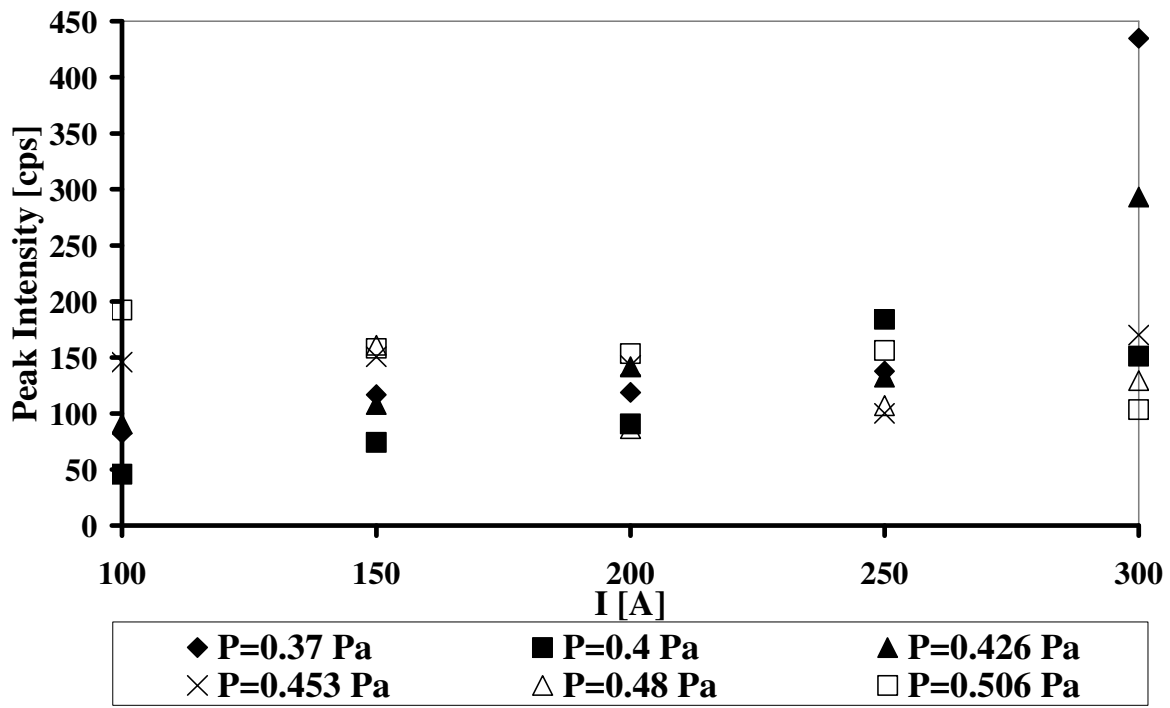

**Figure 3:** Plots of peak intensity of the (002) reflection vs. the arc current I (A), where the oxygen pressure range was 0.37 – 0.51 Pa.

*3.1.4 Grain size*

The grain size can be derived from the XRD reflection width using the Scherer relation:

$$D = \frac{0.94\lambda}{\omega \cdot Cos(\theta)},$$

where D is the grain size in nm, λ is the diffractometer wavelength in nm, $\omega$ is the (002) reflection full width at half maximum after subtracting the instrumental width, and $\theta$ is the diffraction angle. Grain size determined on films deposited with current below 200 A was in the range 10 – 15 nm, and was not correlated with arc current for all pressures used, as shown in Figure 4. Larger spread in grain size was found in films deposited with current ≥ 250 A, 12 – 33 nm. Here, the films with larger grains were deposited at pressure lower than 0.45 Pa.

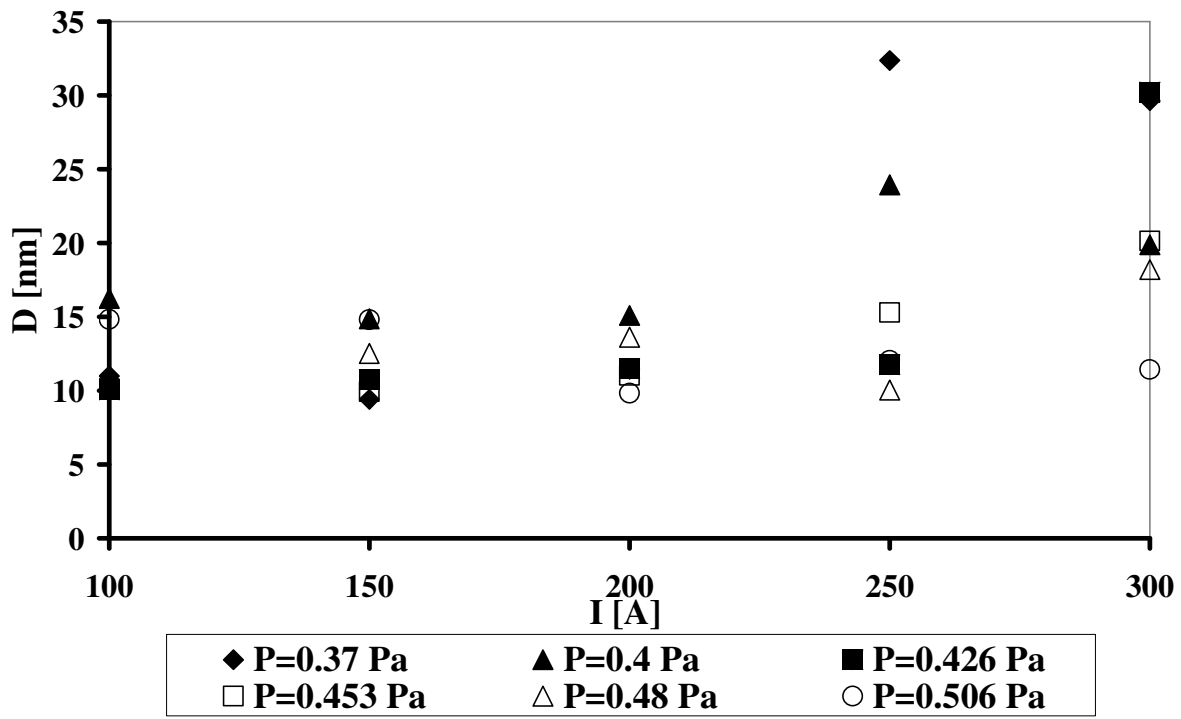

**Figure 4:** Plot of ZnO grain size D (nm) vs. the arc current I (A), where the oxygen pressure range was 0.37 – 0.51 Pa.

*3.1.5 Correlations between structural properties*

The plots of average grain size (D) and elastic stress ($\sigma$) as function of film thickness, h, are presented in Figs. 5a and 5b, respectively, indicating that D was not correlated with the thickness, h, but a correlation between $\sigma$ and h with $R^2 = 0.86$, where R is the correlation coefficient, was noted. Average grain size for all films with h < 500 nm was in the range 10 to 15 nm. The grains in films with thickness > 500 nm were larger, but again no well-defined dependence on thickness was noted. It was noticed that thick films were not necessarily grown with larger grains; however, greater thickness implied higher compressive stress.

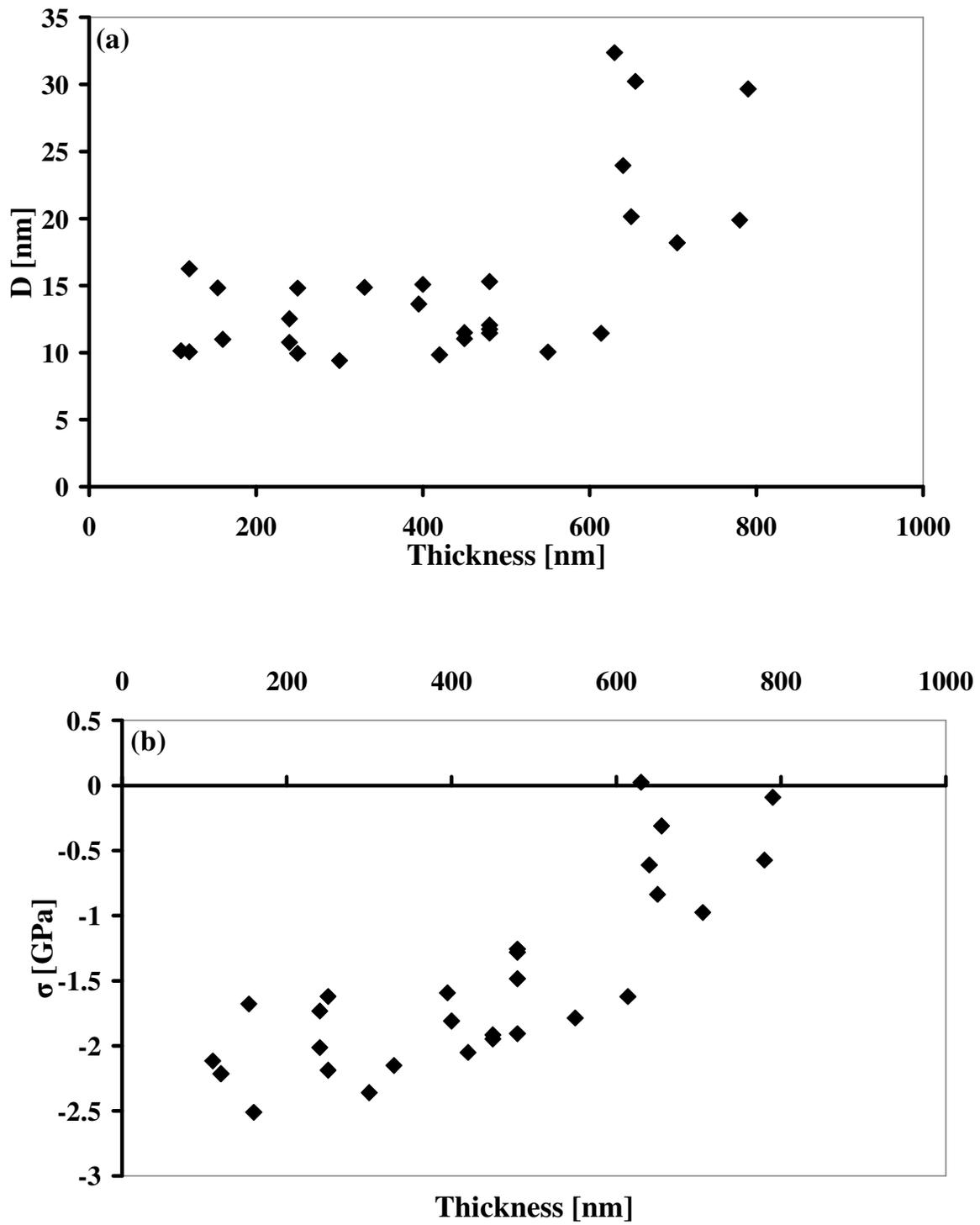

**Figure 5:** (a) Plot of grain size D (nm) vs. film thickness (nm); (b) Plot of film stress σ (GPa) vs. film thickness (nm).

Similarly, no correlation was found between film thickness and the (002) reflection peak intensity. On the other hand, the elastic stress decreased with film thickness. This trend was more pronounced for samples with film thickness > 500 nm.

The average grain size, D, plotted against the elastic stress σ, is shown in Figure 6. Grain size was in the range of 10 – 15 nm for films with σ in the range -2.3 to -1.3 GPa, and no correlation between these parameters was found. In the case of films with σ in the range (-1.3 to +0.4 GPa) a significant linear correlation was established between σ and D, σ = 0.077D-2.32 [GPa], with $R^2 = 96.7\%$.

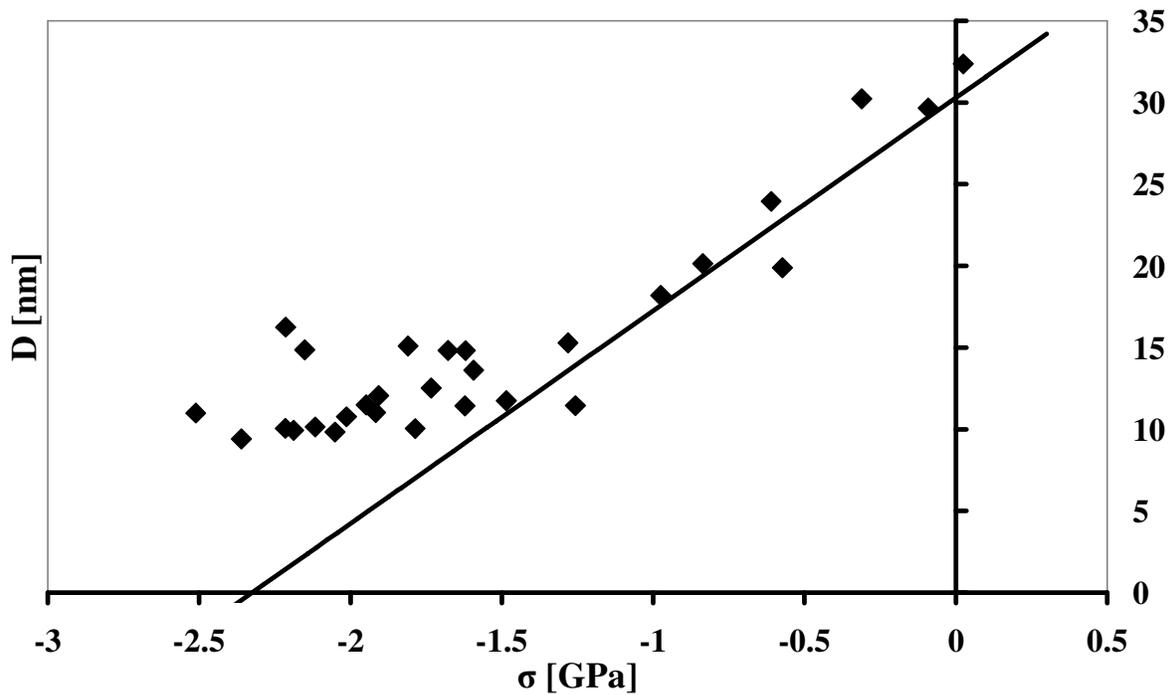

**Figure 6:** Plot of grain size D (nm) vs. the film stress σ (GPa). The solid line is a least squares linear fit.

### 3.2 Compositional analysis

Each sample was analyzed by X-ray photoelectron spectroscopy (XPS). The presence of carbon was observed on film surface in each case, as previously reported [22]. In the bulk of the films, the oxygen-to-zinc ratio ($R_{oz}$) for all films was in the range of 0.68-0.80, with an average of 0.70 and a standard deviation of 0.03 for all samples. No well-defined dependence of $R_{oz}$ on the oxygen pressure or arc current during deposition was observed. The values of $R_{oz}$ of samples deposited with arc currents above 100 A was in the range of 0.68 – 0.72, with an average 0.7 ±0.01. No significant variation in film composition could significantly be associated with pressure fluctuation during the deposition process, as pressure control was very effective, limiting such variation to well below 1%.

A lower atomic concentration of Zn and a higher concentration of oxygen were observed on the surface of the films (typically $R_{OZ}$ ~1.7 to 1.8), as well as concentration of carbon (sometimes up to 50% on the surface). Adsorption of oxygen and carbon from the atmosphere after the deposition could influence the composition of the surface and increase $R_{OZ}$ on the surface. It should be noted however, that the thickness of this surface layer was much less than 5% of the entire film thickness, in all cases.

Oxygen deficiency in thin ZnO film had been reported before. Whangbo et al. [30] observed zinc excess ($R_{OZ} = 0.91$) in ZnO films deposited with a reactive-ionized cluster beam system. Xu et al. [16] reported excess of C on the surface, and $R_{OZ} = 0.61$ in samples deposited using FVA where the substrate was biased by -200 V. Xu et al. [16] assumed that the excess of Zn in those samples resulted from preferred oxygen sputtering by Zn ions. However, Xu et al. [16,17] also reported values of Roz > 1 when the deposition was performed on heated substrates at temperatures of 230 and 430 °C. Their values of Roz increased with substrate temperature, and were 1.02 and 1.1 for films deposited at substrate temperature of 230 °C and 1.16, and1.29 for substrates at 430 °C. They associated the increase in $R_{OZ}$ with substrate temperatures to faster penetration of oxygen into the film, or faster formation of ZnO, resulting in a lower oxygen-sputtering rate. In the next section the effects of the deposition parameters on the microstructure are further discussed.

**3.3 Deposition parameters and film micro-properties**

The two deposition parameters controlled during the experiment were arc current and oxygen pressure. In order to assess their effect it should be noted that an increase in the arc current in a VAD process results in an increase in the plasma flux, but does not result in an increase of the plasma ions energy [32]. Varying the pressure may affect the particle energy by changing the collision frequency between the fast Zn ions and the significantly slower oxygen molecules or atoms, lowering the energy of the Zn ions and the plasma flux. However, assuming a collision cross section of $3.10^{-20}$ m$^2$ and oxygen pressure in the range 0.37 Pa to 0.51 Pa, the mean free path of the Zn ions is estimated to be of the order of the magnetic filter length. Hence, collision effects should be altogether small, and the effect of varying the pressure negligible.

As mentioned above, the Zn vacuum arc produces a plasma jet with ion energy of 37 eV [31]. Hence, the plasma flux to the substrate also heats the substrate, and this could affect the micro-properties of the film. Such heating is proportional to the film thickness [33]. An upper limit to the temperature of the film is obtained by assuming that the kinetic and ionization energies of the deposited Zn ions are completely absorbed in the film, and that there is no heat transfer to the water-cooled sample holder. Such upper limit estimate of the temperature for a 500 nm thick film, after 60 s deposition, is ~80 ºC, and in the case of a 300 nm film thick film ~60 ºC (when room temperature is 25 ºC). The actual increase in the film temperature would be somehow lower, as the water-cooled substrate holder cannot be ignored. It is an open question whether the dependence of some micro-properties on film thickness results from the heating of the film by the plasma flux.

The observed independence of film composition ($R_{OZ}$) on the deposition parameters is in a sense unexpected. Larger plasma fluxes (higher current) and lower oxygen pressure, or lower plasma fluxes (smaller plasma currents) and larger pressure are expected to affect the ratio $R_{oz}$. As this is not the case, it could be argued that under the present experimental conditions, the growth process of the film is controlled by a balancing chemistry and sputtering of oxygen that produce films of approximately constant $R_{oz}$. Further and more detailed investigation is required to corroborate such assumption.

**3.4 Electrical Properties**

*3.4.1 General Observations*

We had recently reported [22] that film thickness was linearly dependent on the pressure, and that the electrical resistivity of the transparent ZnO films was relatively constant as a function of pressure. In the current study it was further observed that film thickness was also linearly dependent on the arc current in the range 100 - 300 A. This relation could be expressed by the formula h = 2.85·I - 156, where h is the film thickness in nm, and I is the current in the range 100 - 300 A (film thickness in the range 90 - 780 nm). The correlation coefficient was $R^2 \sim 0.95$.

Film resistivity, however, was found to be weakly dependent of the current and oxygen pressure, as can be seen from Fig. 7a where the resistivity is plotted against pressure and arc current. Except for one sample, the resistivity of all films is in the range $(0.93 – 4.91) \cdot 10^{-4}$ Ω·m, and no apparent correlation could established between the resistivity and pressure, or resistivity and arc current. It is discussed below that the dispersion of the resistivity in that range results from microstructure effects.

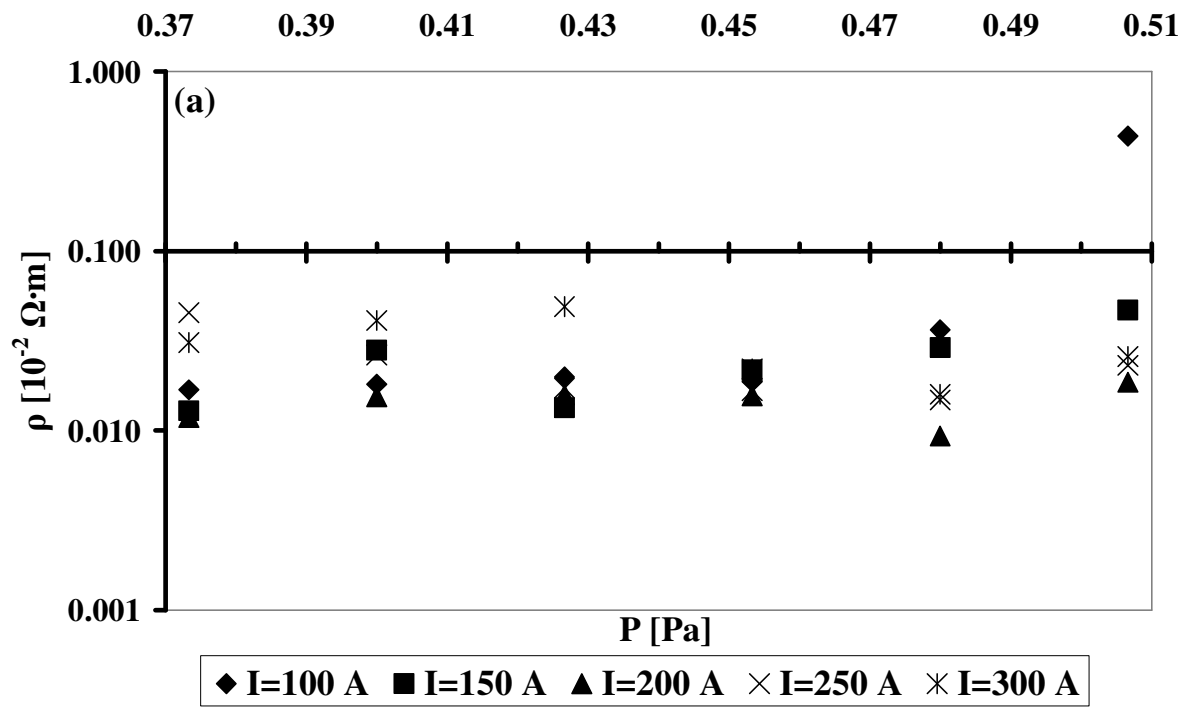
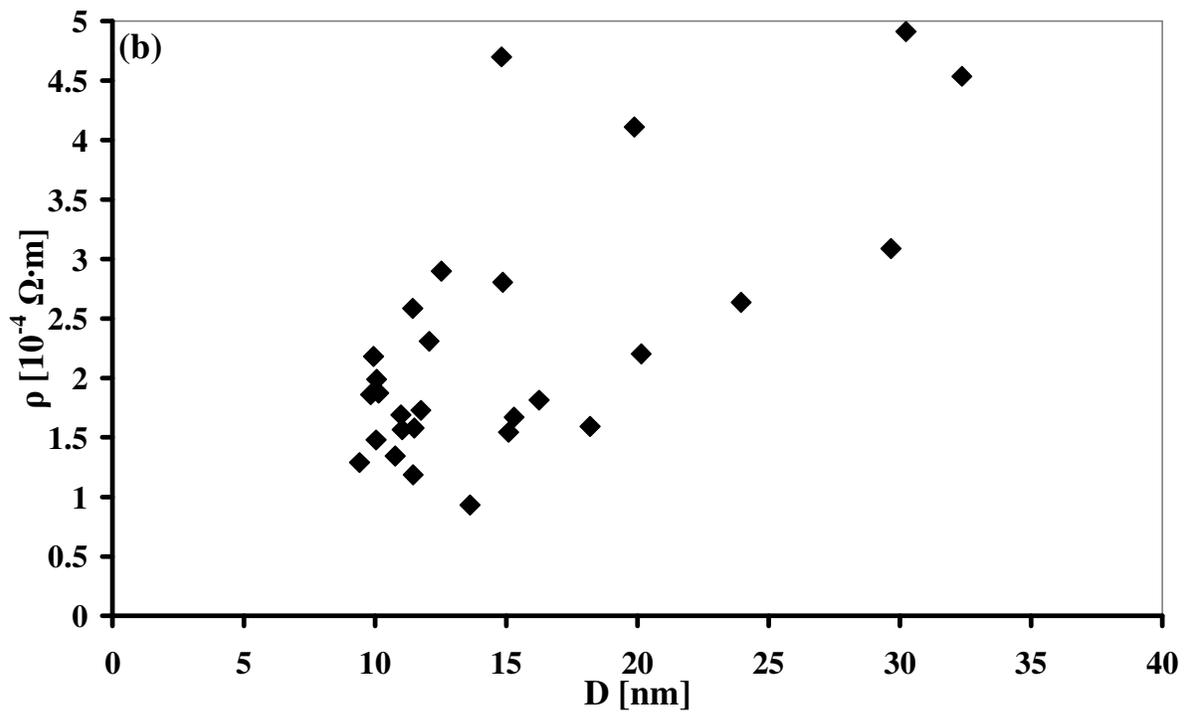

**Figure 7:** Plots of film resistivity ρ (Ω·m) vs. (a) oxygen pressure P (Pa), where the arc current range was 100 – 300 A, and (b) grain size D (nm).

Hall measurement of the resistivity, carrier density, and carrier mobility on a single sample was performed several months after its deposition. The film was deposited with oxygen pressure of 0.43 Pa and 200 A arc current. The data of this analysis could have been affected by the environment during the time interval between deposition and measurement, as the electrical properties of ZnO thin films was shown to vary after deposition when exposed to air [22]. The measured resistivity was $1.12 \cdot 10^{-4}$ $\Omega \cdot$m, in the resistivity range given above. The charge carrier concentration and carrier mobility of this n-type sample were $4.8 \cdot 10^{25}$ m$^{-3}$ and $1.16 \cdot 10^{-3}$ m$^2$/V·s (11.6 cm$^2$/V·s), respectively. These values are in agreement with data presented in literature for untreated ZnO thin films [1,18, 21]. The compiled data for the Hall mobility of thin film ZnO presented by Ellmer [2] indicate that for undoped ZnO polycrystalline films whose resistivity is ~$8 \cdot 10^{-5}$ $\Omega \cdot$m, the carrier density ~$2 \cdot 10^{25}$ m$^{-3}$, and the mobility is ~ $2 \cdot 10^{-3}$ m$^2$/V·s.

*3.4.2 Correlations between the resistivity and micro-properties*

The conduction of electricity in ZnO films at room temperature is characterized by resistivity in the range $10^{-6} – 10^{-2}$ $\Omega \cdot$m [2], in spite of the large energy band gap (~3.2 eV) between the valence and conductance bands. Such relatively low resistance results from the existence of a donor level sufficiently close to the conductance band. The creation of the donor band is usually attributed to zinc excess or oxygen deficiency [34, 35]. The resistivity of the ZnO film, however, is not determined only by the carrier density; it is also a function of the ZnO film polycrystalline structure, which could affect the carrier density and the mobility of the electrons in the material. In the present experiment, the resistivity was partially correlated only with grain size D, but not with oxygen pressure or arc current. Two factors could lead to the observed spread of film resistivity: a spread in Roz and microstructure variation resulting from variation in the deposition parameters. However, as the standard deviation of Roz was small, < 2% for most samples, other characteristics of the film, not Roz, should have some effect in determining the resistivity, as is evident by the partial correlation between the resistivity and average grain size D seen in Fig. 7b, where the data is scattered, yet a trend could be identified.

The resistivity of polycrystalline ZnO can be described by two electrical resistivity models, also appropriate to model the resistivity of other (doped) polycrystalline TCO's. The two models differ in the basic mechanism responsible for the electrical resistance. In TCO's with high carrier density ($> 10^{26}$ m$^{-3}$), the resistivity is based on carrier scattering by ionized impurities (intrinsic lattice defects or extrinsic dopants) [2, 36, 37]. In polycrystalline TCO's with lower carrier density ($< 10^{26}$ m$^{-3}$), the resistivity model of Seto [38] and Bruneaux [39] is applicable, and the resistivity is mostly due to the grain barrier electron trapping. Seto [38] showed that his model agreed closely with the resistance of phosphorous-doped polycrystalline Si, while Bruneaux [39] applied it to fluorine-doped tin dioxide films. In the present experiment the resistivity of grains with D in the range 10 – 20 nm was (1- 3)·10$^{-4}$ Ω·m, implying according to the data in Ref. 2 carrier density $\leq 10^{25}$ m$^{-3}$ (as also supported by the data from the Hall measurement mentioned above), hence, the resistivity is assumed to be determined by carrier trapping and grain boundary potential barrier, neglecting the bulk resistance of the grains, in accordance to the model of Seto. [38] According to the grain boundary model, in order for the resistivity to grow with D the area density of traps on the grain boundary (Qt) should also grow with D and with the density of the impurity ions (N), i.e. excess of Zn ions. However, as the variation of O/Zn ratio over the range is smaller than 8%, it is reasonable to assume that the change in Qt was only weakly dependent on N. Hence, the increase of ρ with D probably implies an increase of Qt. A direct determination of Qt is required to support the application of the grain boundary model for the connection between ρ and D. Similar report on the resistivity of oxygen deficient ZnO films were reported by Whangbo et al. [30], who measured resistivities in the range ($10^6$ – $10^{-4}$)·Ω·m.

Large-grained films were more resistive. No correlation was found between the resistivity and X-ray (002) reflection peak intensity, the latter representing the film crystalline quality. This result agrees well with the observation that the intensity of the (002) reflection was not correlated with D while $\rho$ was. It also supports the assumption that the resistivity is determined by the boundary effect model, and the quality of the grains might have only a secondary effect on the resistivity.

**3.5 Optical analysis**

*3.5.1 Optical Transmission*

Film thickness affects their optical transmission, and in order to eliminate the thickness factor, the *extinction coefficient (α)*, which is a function of the wavelength, was determined from the transmission data, according to the expression:

$$T_r = e^{-\alpha \cdot d},$$

where $T_r$ is the transmission of a ZnO film with thickness d.. Thus, small values of α indicate high transmission, or large e-folding thickness.

The variation of α with arc current is presented in Figures 8a-8c. In all cases, ZnO films deposited with higher arc current had larger α throughout the observed spectrum. As function of wavelength, α of all films deposited with arc current<250 A decreased monotonically with wavelength in the whole observed range. However, α of films deposited with arc current >200 A and pressure< 0.43 Pa started to decrease for wavelength > 430 or 440 nm, after a slight increase in the range 380-440. The position of the maximum in α varied with arc current. The lowest α–value was observed for a film deposited with 100 A arc current, around $6 \cdot 10^{-4}$ nm$^{-1}$, resulting with a ~90% transmission for a film thick 210 nm. The values of the parameter α, derived for samples deposited with pressure of 0.4 Pa are shown in Fig. 8b. Here, the values of α for films deposited with arc currents in the range 100 – 200 A were very close, having practically the same α for λ > 540 nm. The values of α of samples deposited with arc current larger than 200 A and with P=0.4 Pa, though higher than the α of samples deposited at lower currents, decreased monotonically with wavelength, approaching that of samples deposited at P = 0.37 Pa. The extinction coefficient of samples deposited with the higher pressure (0.5 Pa) depended only weakly on the current (Figure 8c), and had a significantly lower α.

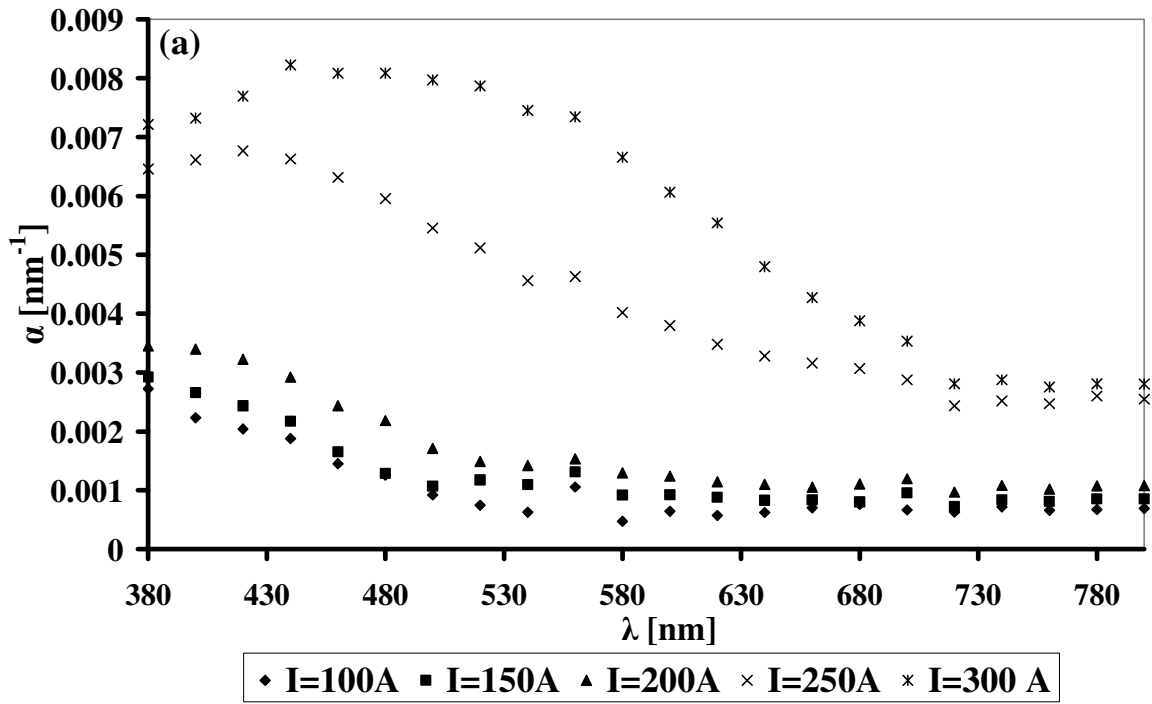

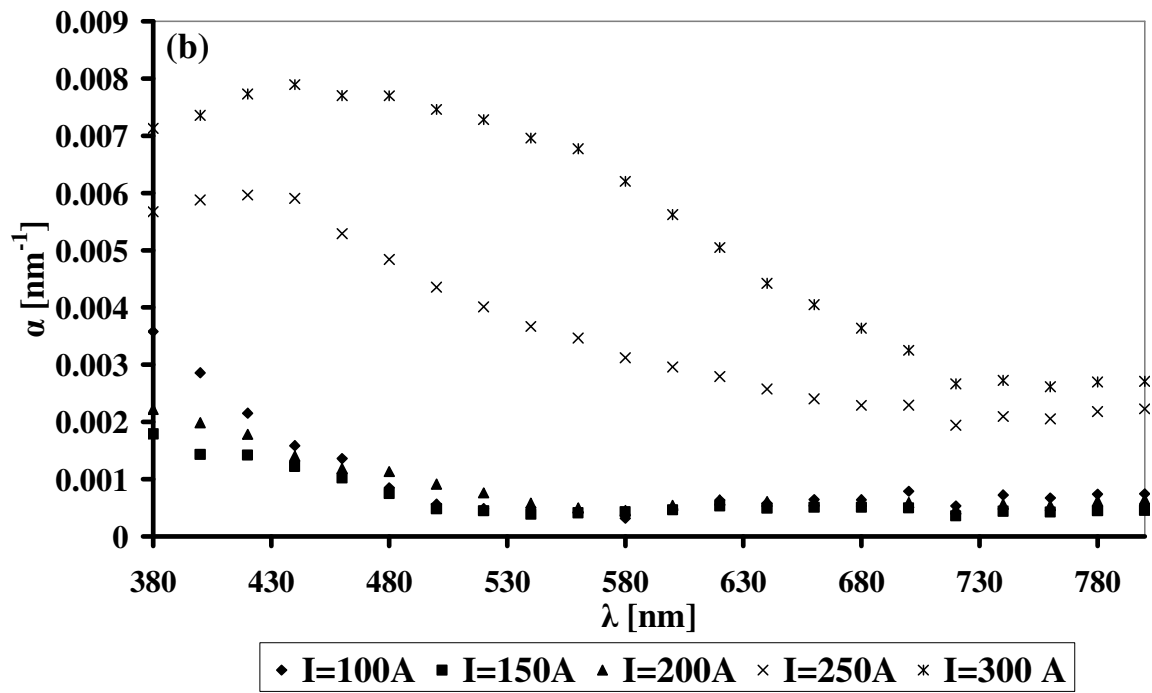

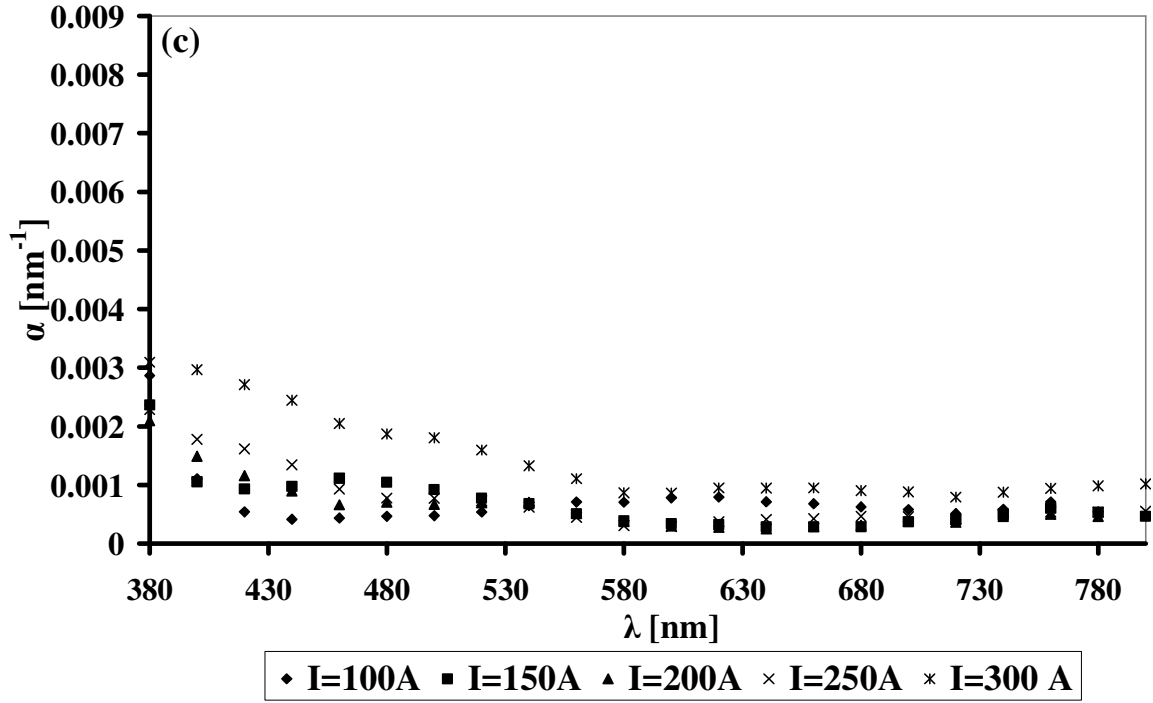

**Figure 8:** Plot of the extinction parameter α (nm$^{-1}$) vs. wavelength (λ), for the following oxygen pressures: (a) P=0.37 Pa, (b) P=0.4 Pa, and (c) 0.51 Pa.

*3.5.2 Correlations between the optical transmission and micro-properties*

No well-defined correlation was found between the film optical transmission, i.e. α, and film composition, or between α and the (002) reflection peak intensity. Practically no correlation was found between α and the grains size, where above D = 16 nm the films had very low transmission (T$_r$ < 0.1). This was to be expected, as these were the thicker films (Fig. 5a). The observation that α was weakly correlated with grain size indicated negligible absorption and scattering by the grains. Mie's scattering and absorbing theory applies to a medium containing dielectric and conducting spheres, but should also provide a reasonable approximation to non-spherical grains provided the dimensions of the grain are much smaller than the wavelength [40]. The complex index of refraction of ZnO in the VIS region is approximately 2+ik, where k~ 0.05 – 0.2, implying that the parameter δ and β defined by van de Hulst are ~1 and ~3º – 10º, respectively [41]. As β < 15° and the ratio $2\pi D/\lambda \cong 0.25$, hence the absorption and scattering cross section are $<< \pi D^2/10$, and the mean free path of a photon will be at least 50D. As film thickness is in the range (6 – 15)·D, the effect of the grains on the transmission would be negligible.

## *4. Conclusions*

The variation of arc current and oxygen pressure of a FVAD system affected in a complex manner the electrical conductivity, optical transmission, chemical composition, and structure of thin ZnO film (70 to 780 nm thick), which were deposited for 60 s. In most cases the experimental data did not indicate an ordered relation between deposition parameters and film characteristics. The deposition that was performed with oxygen pressure varying the range 0.37 – 0.51 Pa and arc current of 100 – 300 A, resulted in polycrystalline, hexagonal c-oriented TCO films, consisting of 10 –35 nm crystalline grains, with internal compressive stress in the range –2.5 to 0 GPA.

The variation of the deposition parameters affected weakly the film chemical composition, which always showed zinc excess in the bulk. The overall oxygen-to-zinc atomic concentration ratio was in the range of 0.68-0.80, however, in most cases the ratio was in the range 0.68-0.72. No definite correlations were found between the structural properties and the composition. Film thickness depended linearly on deposition time, arc current, and oxygen pressure.

The electrical resistivity ρ of the transparent films $((1 – 5)\cdot 10^{-4}$ Ω·m) was correlated with film grain size and depended on arc current only for arc current ≥ 250A. No distinct dependence of ρ on the oxygen pressure was found. The dependence of the extinction coefficient α on the deposition parameters did not show a distinct trend, but the optical transmission of the films, which depends on both α and film thickness, increased with oxygen pressure and decreased with arc current, reaching a maximal value up to 97% over the visual and near-IR ranges of the spectrum, for films of ~200 nm thickness.

We conclude that although the characteristics of thin ZnO films deposited with FVAD system could be partially affected by adjusting the deposition parameters, the degree of control was limited.

Further study is required including investigating the effects of additional deposition parameters, e.g. substrate temperature and electrical bias, to determine how to achieve a better control of film properties by changing the deposition parameters. In the present case, the best combination of lowest resistivity and lowest extinction coefficient was obtained for a film deposited with I = 200 A and pressure of 0.5 Pa, with $\rho = 10^{-4}$ $\Omega \cdot$m, and $\alpha = 0.0004$ nm$^{-1}$ at wavelength of 600 nm.

*Acknowledgements*

The authors thank Dr. L. Burstein and Dr. Yu Rosenberg for the XPS and XRD measurements. This research was partially supported by a grant from Tel Aviv University, The Gordon Center for Energy Studies.